\documentclass[epj]{svjour}
\usepackage{amsmath}
\usepackage{amssymb}
\usepackage{amsbsy}
\usepackage{graphicx}
\usepackage{color}

\begin{document}
\title{Anomalous force-velocity relation of driven inertial tracers in steady laminar flows}
\author{F. Cecconi \inst{1} \and A. Puglisi \inst{1} \and A. Sarracino\inst{1} \and A. Vulpiani \inst{2,3}}                     % Do not remove
\institute{CNR-ISC and Dipartimento di Fisica, Sapienza Universit\`a di Roma, p.le A. Moro
2, 00185 Roma, Italy \and Dipartimento di Fisica, Sapienza Universit\`a di Roma,  and  CNR-ISC, p.le A. Moro
2, 00185 Roma, Italy \and Centro Interdisciplinare B. Segre, Accademia dei Lincei}
\date{Received: date / Revised version: date}
\abstract{We study the nonlinear response to an external force of an
  inertial tracer advected by a two-dimensional incompressible laminar
  flow and subject to thermal noise. In addition to the driving
  external field $F$, the main parameters in the system are the noise
  amplitude $D_0$ and the characteristic Stokes time $\tau$ of the
  tracer. The relation velocity vs force shows interesting effects,
  such as negative differential mobility (NDM), namely a non-monotonic
  behavior of the tracer velocity as a function of the applied force,
  and absolute negative mobility (ANM), i.e. a net motion against the
  bias. By extensive numerical simulations, we investigate the phase
  chart in the parameter space of the model, $(\tau,D_0)$, identifying
  the regions where NDM, ANM and more common monotonic behaviors of the
  force-velocity curve are observed.
 % end of PACS codes
} %end of abstract
\maketitle
%
%%%%%%%%%%%%%%%%%%%%%%%%%%%%%%%%%%%%%%%%%%%%%%%%%%%%%%%%%%%%%%%
\section{Introduction}
\label{sec:intro}
%%%%%%%%%%%%%%%%%%%%%%%%%%%%%%%%%%%%%%%%%%%%%%%%%%%%%%%%%%%%%%%
The study of the transport properties of inertial particles in fluids
takes on a great importance in several fields, in engineering as well
as in natural occurring settings: typical examples are pollutants and
aerosols dispersion in the atmosphere and oceans \cite{Pollutant},
optimization of mixing efficiency in different contexts, or the study
of chemical \cite{Chem}, biological \cite{CEN13} or physical
interaction \cite{MEH16}, with applications to the time scales of
rain~\cite{F02}, in the sedimentation speed under
gravity~\cite{LCLV08}, or to the planetesimal formation in the early
Solar System~\cite{B99}.

Under external perturbations, the dynamics of tracer particles can be
significantly modified, resulting in different behaviors which are not
easily predicted from the properties of the unperturbed dynamics.  In
order to relate response functions and fluctuations in non-equilibrium
conditions, relevant in the aforementioned cases, generalizations of
the standard fluctuation-dissipation theorem have been derived in
recent years~\cite{CLZ07,MPRV08,LCSZ08,LCSZ08b,seifert,BM13}.
These approaches are generally valid in the small forcing limit, and a
central problem remains the study of the motion in the presence of an
external driving field $F$, in the nonlinear regime.  In this case,
the particles reach a stationary state characterized by a finite
average velocity $v$ which non-trivially depends on the system
parameters.  The main point is then to understand the force-velocity
relation $v(F)$, which contains relevant information on the
system.  These curves are strongly affected by the
nature of the tracer/fluid interaction, and can show surprising
nonlinear behaviors.

An example of such effects is provided by the so-called negative
differential mobility (NDM), which means that $v(F)$, after a linear
increase with the applied force, can show a non-monotonic behavior,
attaining a (local) maximum for a given value of the
force~\cite{royce,LF13,BM14,BIOSV14,BSV15,BIOSV16}.  For larger
intensity of the bias the velocity can show different behaviors, such
as saturation or asymptotic linear growth, depending on the considered
model.  In other situations, one can also observe a more surprising
feature: an \emph{absolute} negative mobility (ANM), where the
particle travels on average against the external
field~\cite{RERDRA05,KMHLT06,MKTLH07,ERAR10}.  In general, these
non-linear behaviors are due to trapping mechanisms in the system,
which lead to dynamical conditions such that an increase of the
applied force can result in an increase of the trapping time, and,
consequently, to a slowing down of the average tracer dynamics.
Depending on the specific model, trapping can be due to the
interaction of the tracer with the surrounding particles, to
frustration in the system, to geometric constraints or to the coupling
with underlying velocity fields.

Here we study the response to an external bias of inertial particles
advected by steady (incompressible) cellular flows, in the nonlinear
regime.  In these systems, the presence of inertia induces a
non-trivial deviation of the particle motion from the flow of the
underlying fluid: the particles can remain trapped in regions close to
upstream lines, yielding a slowing down of the dynamics, as observed
in the context of gravitational settling~\cite{M87,RJM95,F97,A07}, and
typically leading to the phenomenon of preferential
concentration~\cite{BBCLMT07,CCLT08}.  In a recent work~\cite{SCPV16},
we have considered a model where, in addition to the cellular flow and
the external force, the inertial tracer is subject to the action of a
microscopic (thermal) noise.  We have shown that a rich nonlinear
behavior for the average particle velocity can be observed, as a
function of the applied force, featuring both NDM and ANM.  The latter
was never observed in the standard systems studied in the literature.
Here we present an extensive numerical investigation of the system
studied in~\cite{SCPV16}, exploring a wide range of the model
parameters and reconstructing the phase chart, where regions of NDM
and ANM are identified.

%%%%%%%%%%%%%%%%%%%%%%%%%%%%%%%%%%%%%%%%%%%%%%%%%%%%%%%%%%%%%%%
\section{The model}
\label{sec:model}
\begin{figure}[!t]
\includegraphics[angle=-90,width=.8\columnwidth,clip=true]{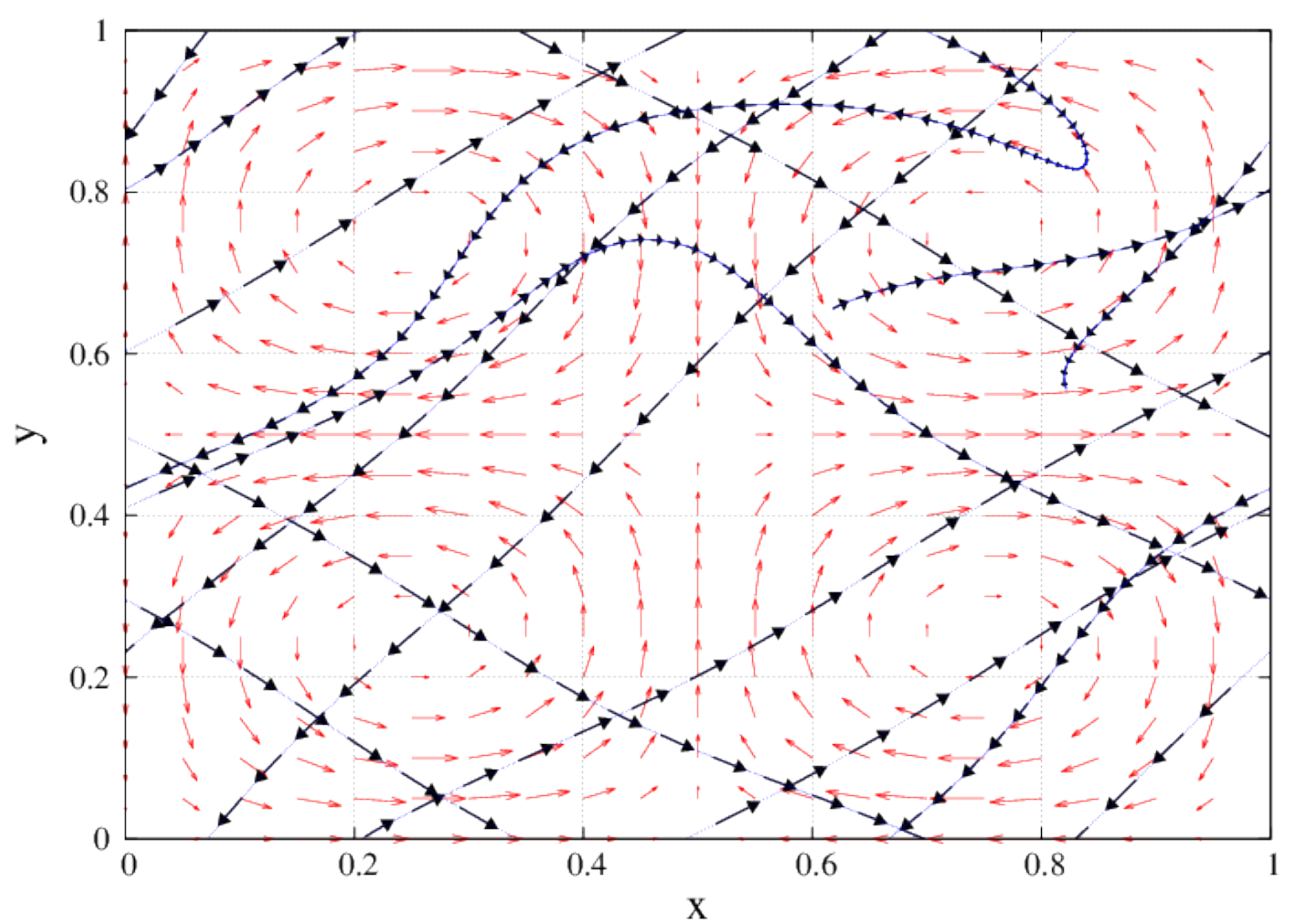}
\caption{Sample trajectory of an inertial particle (black arrows) advected by
  the underlying velocity field (red arrows), in the presence of
  external force and thermal noise, for parameters $\tau=10$, $F=10^{-2}$,
  $D_0=10^{-5}$.}
\label{fig:model}
\end{figure}

%%%%%%%%%%%%%%%%%%%%%%%%%%%%%%%%%%%%%%%%%%%%%%%%%%%%%%%%%%%%%%%
The dynamics of the inertial tracer in two dimensions, with
spatial coordinates $(x,y)$ and velocities $(v_x,v_y)$, is described
by the following equations
\begin{eqnarray}
\dot{x}&=&v_x, \qquad \dot{y}=v_y \label{eq1} \\ 
\dot{v}_x&=&-\frac{1}{\tau}(v_x-U_x)+F +\sqrt{2D_0}\xi_x \label{eq2}, \\
\dot{v}_y&=&-\frac{1}{\tau}(v_y-U_y)+ \sqrt{2D_0}\xi_y, \label{eq3}
\label{model}
\end{eqnarray}
where $\mathbf U = (U_x,U_y)$ is 
a divergenceless cellular flow
defined by a stream-function $\psi$ as:
\begin{equation} 
U_x=\frac{\partial \psi(x,y)}{\partial y}, \qquad U_y=-
\frac{\partial \psi(x,y)}{\partial x}\;. 
\label{eq:psi}
\end{equation}
Here $\tau$ is the Stokes time, $F$ the external force, and
\begin{equation}
\psi(x,y) = LU_0/2\pi \sin(2\pi x/L)\sin(2\pi y/L). 
\end{equation}
$\xi_x$ and $\xi_y$ are uncorrelated white noises with zero mean and
unitary variance. A pictorial representation of the field (red arrows)
and of a tracer's trajectory (black arrows) is reported in
Fig.\ref{fig:model}.  Measuring length and time in units of $L$ and
$L/U_0$ respectively, and setting therefore $U_0=1$ and $L=1$, the
typical time scale of the flow becomes $\tau^*=L/U_0=1$.  Let us
anticipate that the time scales ratio $\tau/\tau^*$ will play a
central role in the behavior of the system.  Another important
parameter of our model is the microscopic thermal noise with
diffusivity $D_0$, which guarantees ergodicity and can be expressed in
terms of the temperature $T$ of the environment by the relation
$D_0=T/\tau$.

In the following we will focus on the force-velocity relation, namely
on the behavior of the stationary velocity $\langle v_x \rangle = \tau
F + \langle U_x[x(t),y(t)] \rangle$, where $\langle\cdots\rangle$
denotes averages over initial conditions and noise realizations.  The
numerical integration of the dynamical equations of the model is
performed with a second-order Runge-Kutta algorithm~\cite{honey}, with a time step
$\Delta t=10^{-2}$. Numerical results shown in the figures are averaged over
about $10^{4}$ realizations.  In our study, we will consider different
regimes of the time scale ratio $\tau/\tau^*$, exploring the behavior
of the force-velocity relation $\langle v_x\rangle(F)$ by varying the
microscopic diffusivity $D_0$.
%--------------------------- Fig.1 --------------------------------
%-----------------------------------------------------------------

%%%%%%%%%%%%%%%%%%%%%%%%%%%%%%%%%%%%%%%%%%%%%%%%%%%%%%%%%%%%%%%%%%%
\section{Force-velocity relation}
%%%%%%%%%%%%%%%%%%%%%%%%%%%%%%%%%%%%%%%%%%%%%%%%%%%%%%%%%%%%%%%%%%%

%==================================================================
\subsection{Small Stokes time}
\begin{figure}[!t]
\includegraphics[width=.8\columnwidth,clip=true]{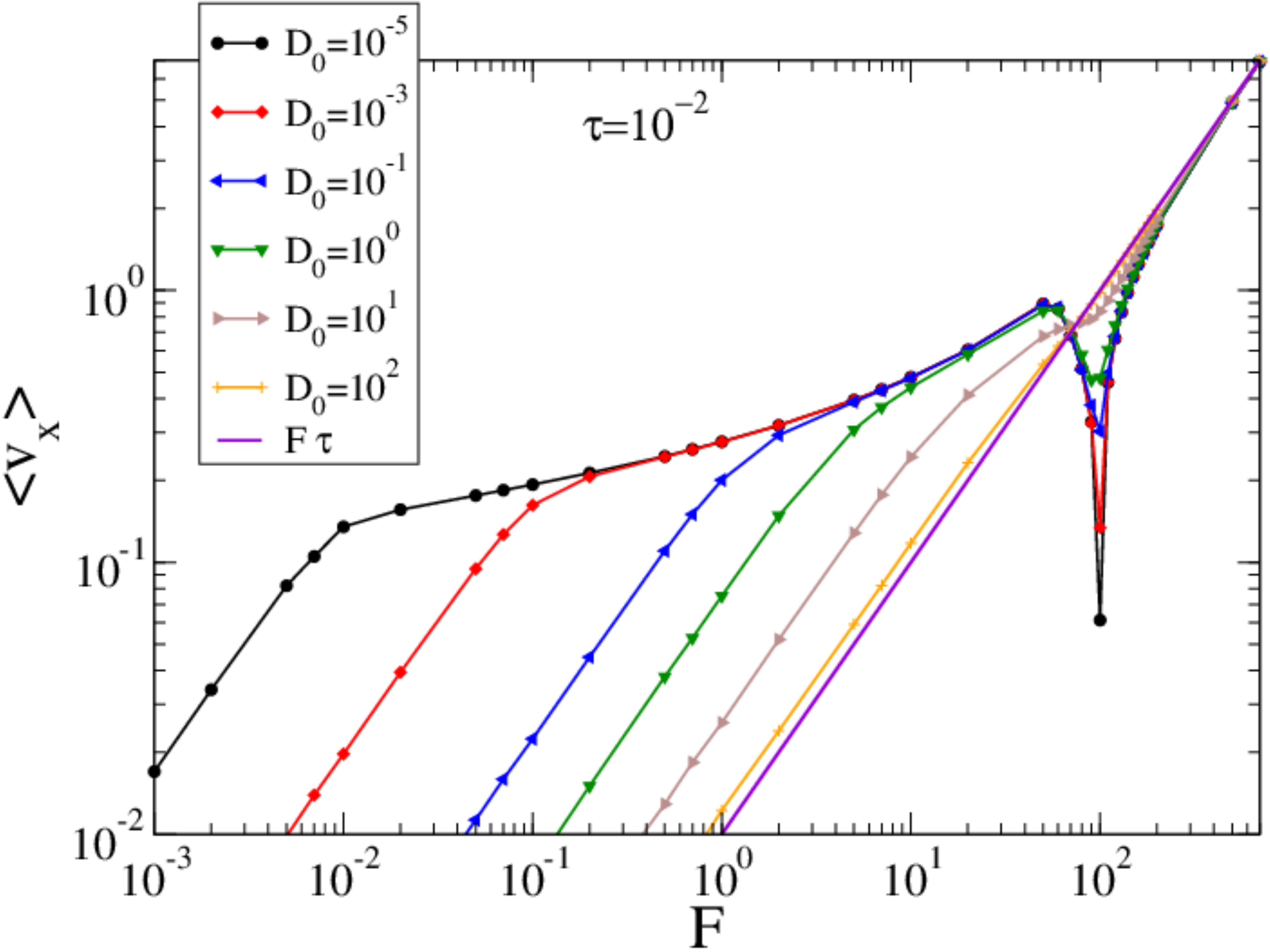}
\includegraphics[width=.8\columnwidth,clip=true]{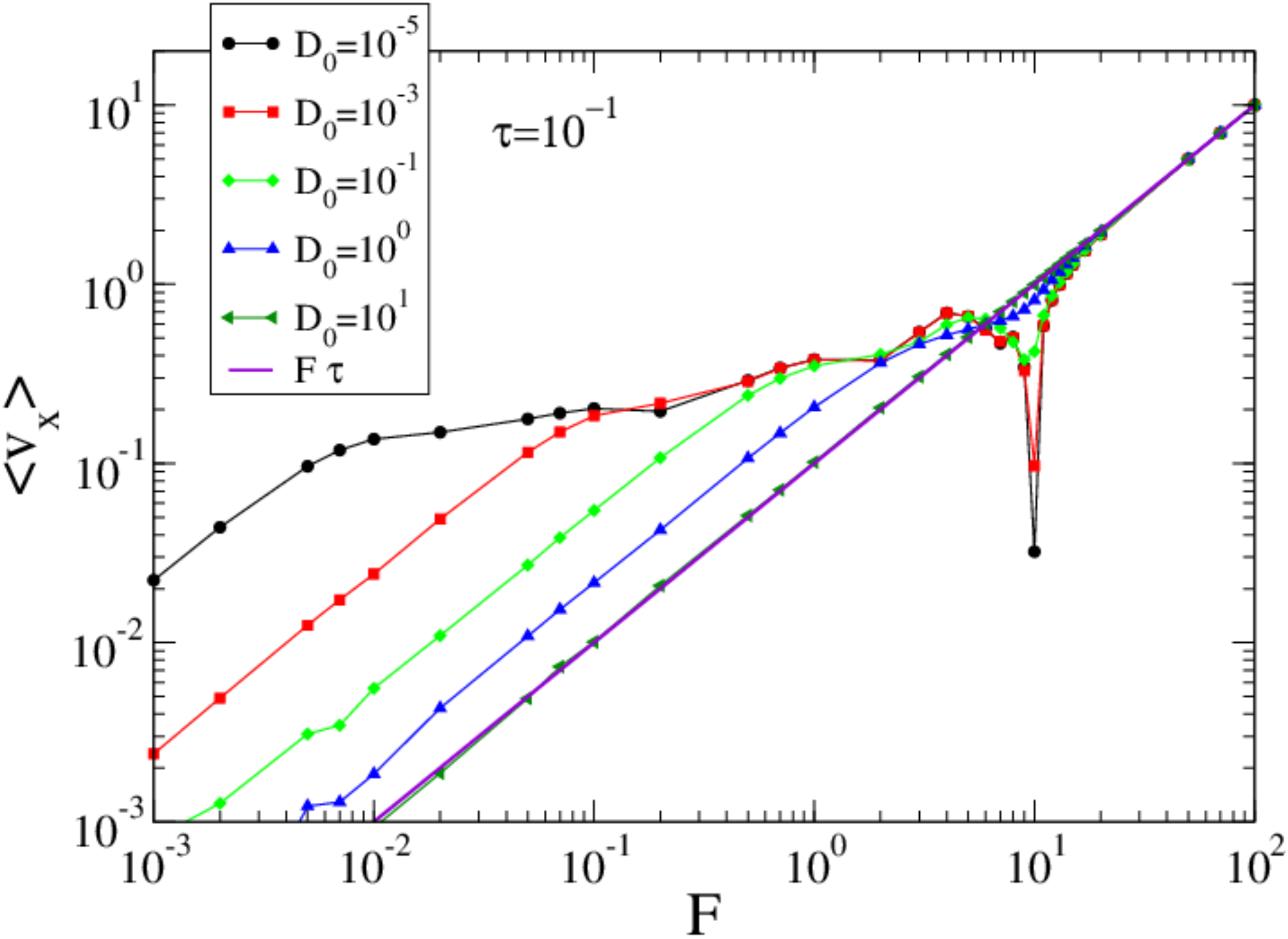}
\caption{Plot of $\langle v_x\rangle$ versus $F$ several values of $D_0$. 
Top panel refers to $\tau=10^{-2}$, while bottom panel refers to $\tau=10^{-1}$ 
Note the abrupt drops of the curves at the value $F=F^*= 1/\tau$.}
\label{fig:lowtau-VF}
\end{figure}

%==================================================================
First, we consider the case $\tau\ll\tau^*$. 
Fig.\ref{fig:lowtau-VF} shows $\langle v_x \rangle(F)$ for 
$\tau=10^{-2},10^{-1}$ and different values of $D_0$.  
Two linear regimes at small and large forces are well apparent. 
In the latter case, the simple behavior $\langle v_x\rangle(F)=\tau F$ is 
recovered. 
This is expected because at large forces the underlying velocity field is irrelevant. 
More surprisingly, in the range of intermediate forces a non-trivial nonlinear 
behavior takes place.  
%-----------------------------FIG.2---------------------------------------
%-------------------------------------------------------------------------
In particular, a non-monotonic behavior, corresponding to NDM, can be observed for 
small values of $D_0$. 
Note that the critical force value $F^*$ where the abrupt drop of velocity
occurs is independent of the noise amplitude and scales with 
$\tau$ as $F^* = \tau^{-1}$.  
As $D_0$ is increased, the effect of the velocity field is averaged out and the
force-velocity curve displays a simple monotonic behavior.

\begin{figure}[!th]
\centering
\includegraphics[width=.48\columnwidth,angle=-90,clip=true]{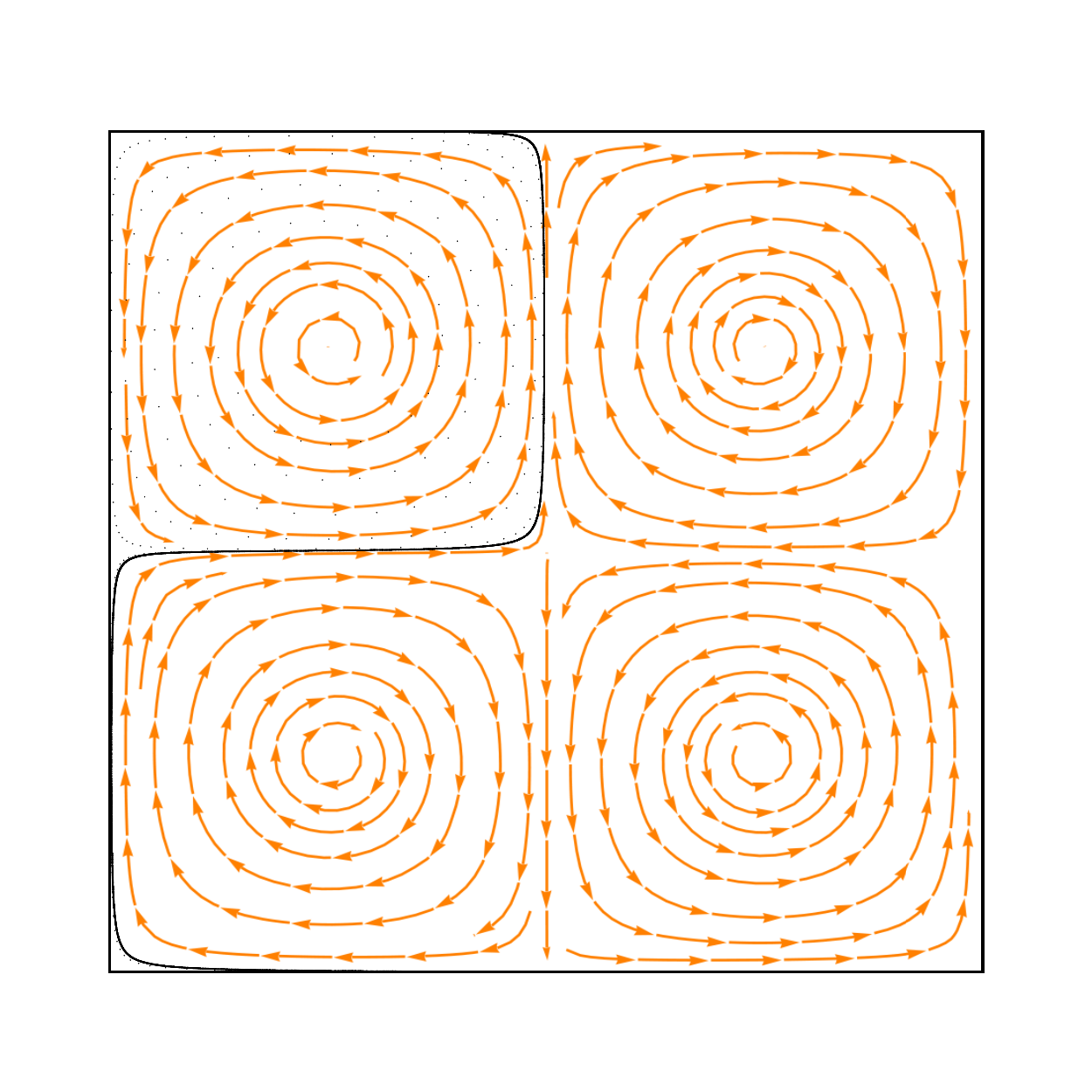}
\includegraphics[width=.48\columnwidth,angle=-90,clip=true]{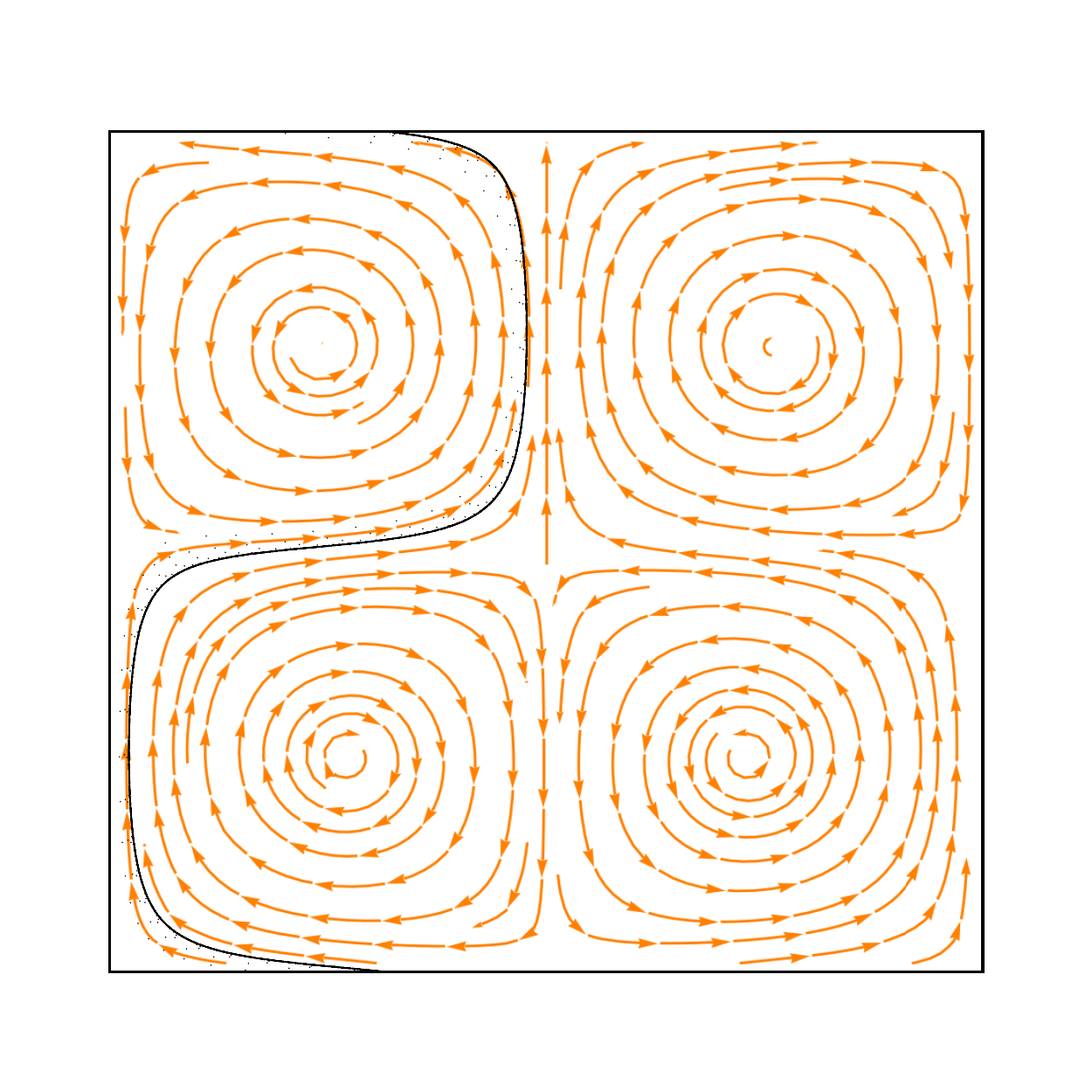}
\includegraphics[width=.48\columnwidth,angle=-90,clip=true]{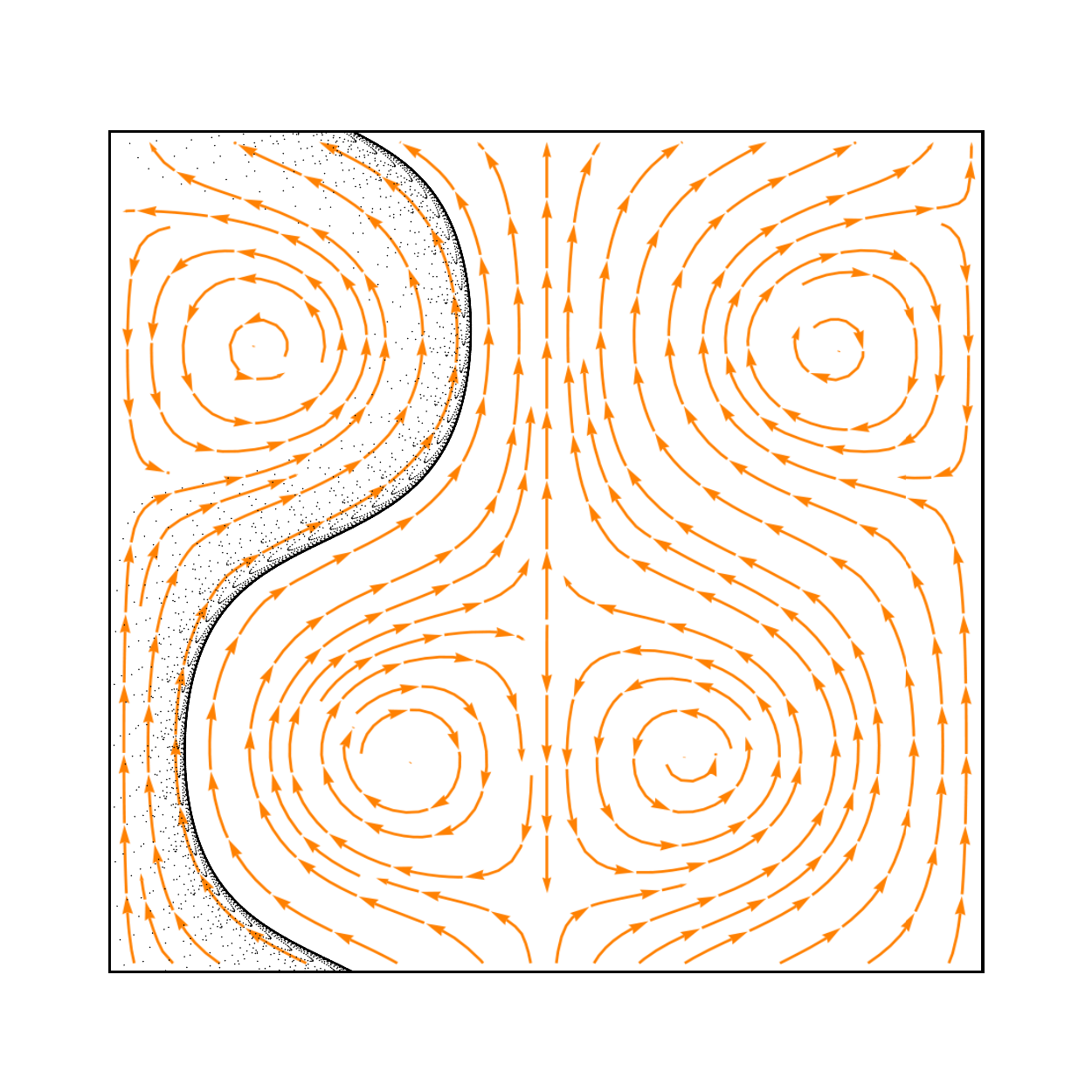}
\includegraphics[width=.48\columnwidth,angle=-90,clip=true]{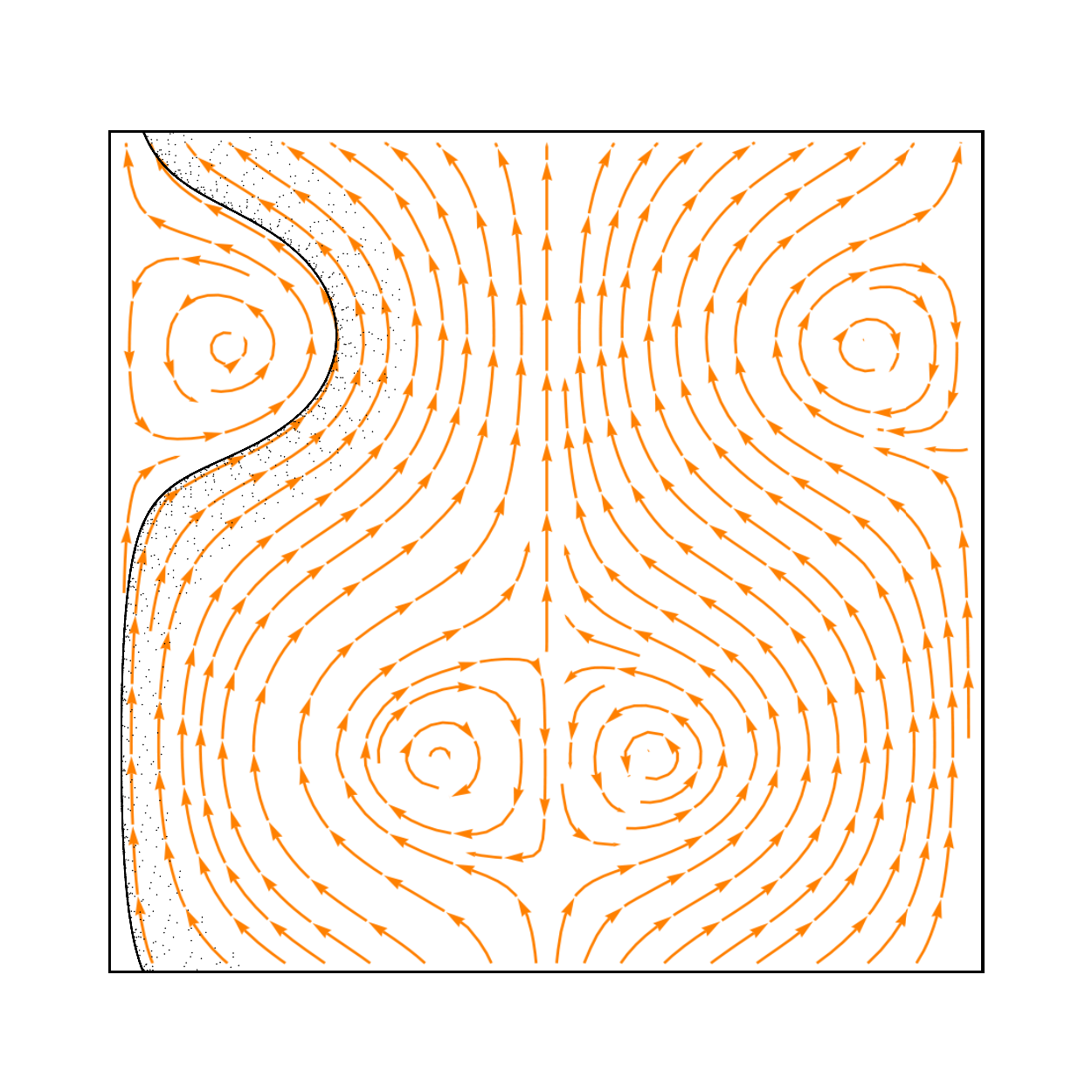}
\includegraphics[width=.48\columnwidth,angle=-90,clip=true]{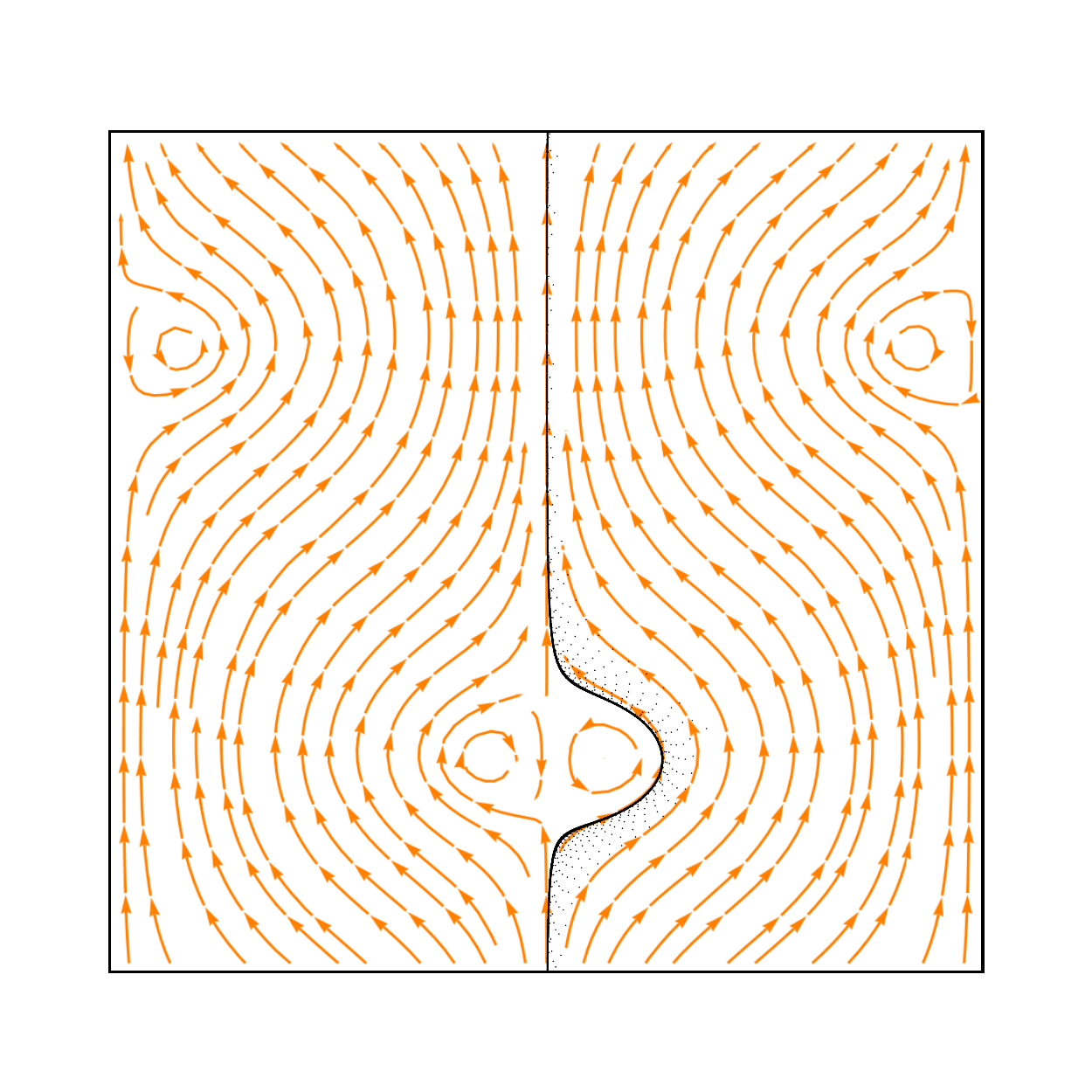}
\includegraphics[width=.48\columnwidth,angle=-90,clip=true]{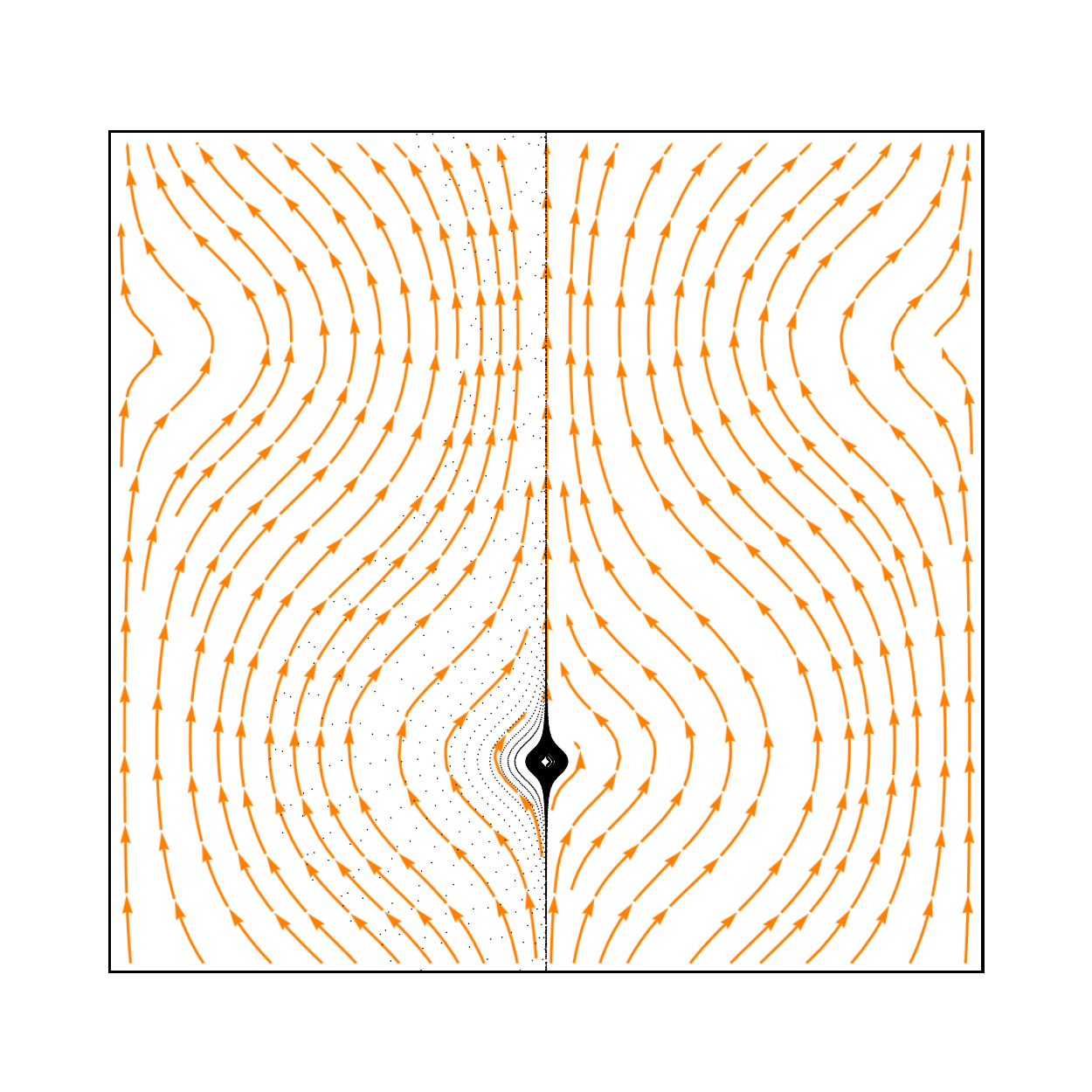}
\caption{Black dots represent trajectories of the full equations of
  motions (Eqs.~\eqref{eq1}-\eqref{eq3}) for driven inertial tracer in
  the case $\tau=0.01$ and $D_0=10^{-5}$, for some values of the force
  $F=1,10,50,70,90,100$ from left to right and from top to
  bottom. Orange arrows represent the streamlines of the effective
  flow in Eqs.~\eqref{eq:slow1}-\eqref{eq:slow2}, obtained
  implementing a geometric singular perturbation approach in the
  Stokes time $\tau$ from the system with $\tau=0$.}
\label{fig:traj}
\end{figure}

This scenario is illustrated by Fig.\ref{fig:traj}, showing
trajectories of the tracer in the case $\tau=0.01$ and $D_0=10^{-5}$,
for some values of the force $F\in [1,100]$ along with the streamlines
of the effective flow obtained implementing a geometric singular
perturbation approach in the Stokes time $\tau$ from the system with
$\tau=0$.  According to Fenichel \cite{FEN79}, for small $\tau$,
particle trajectories of the system are attracted by a two-dimensional
slow manifold and the equations of motion along the manifold can be
formally written as a perturbation series
\begin{eqnarray}
\dot{x}&=& U_x(x,y) + F\tau + \sum_{n} \tau^{n}\;h_{n}(x,y) \label{eq:slow1}\\ 
\dot{y}&=& U_y(x,y)         + \sum_{n} \tau^{n}\;k_{n}(x,y) \label{eq:slow2}
\end{eqnarray}
where the terms $\{h_n,k_n\}$ are explicitly derived in
Refs.~\cite{RJM95,BUR99}.  Eqs.~(\ref{eq:slow1},\ref{eq:slow2}), also
referred to as {\em inertial equations}, have the advantage of
reducing the dimensionality of the system from four to two, yet
catching the correct asymptotic behavior of full-system trajectories.
The drawback lies in the obvious technical difficulties to control the
convergence and in dealing with truncation errors. Let us note that
such a limit is singular and, since now we are in two dimensions,
chaos cannot exist.  In our case, retaining only the 4-th order terms
in $\tau$ in Eqs.~(\ref{eq:slow1},\ref{eq:slow2}) is sufficient to
discuss the qualitative features of the trajectories on the slow
manifold, see Fig.\ref{fig:traj}.  At small forces ($F=1,10,50$), the
asymptotic trajectories (black dots) accumulate along the main
streamlines of the effective velocity field, following the external
force.  For larger values ($F=70,90,100$), one observes an interesting
phenomenon: the trajectories move towards upstream regions with
respect to the external force, so that the inertial tracers slow their
motion, decreasing the stationary average velocity.
%-----------------------------FIG.3------------------------------------
%-----------------------------------------------------------------------

%==================================================================
\subsection{Large Stokes time}
\begin{figure}[!ht]
\includegraphics[width=.8\columnwidth,clip=true]{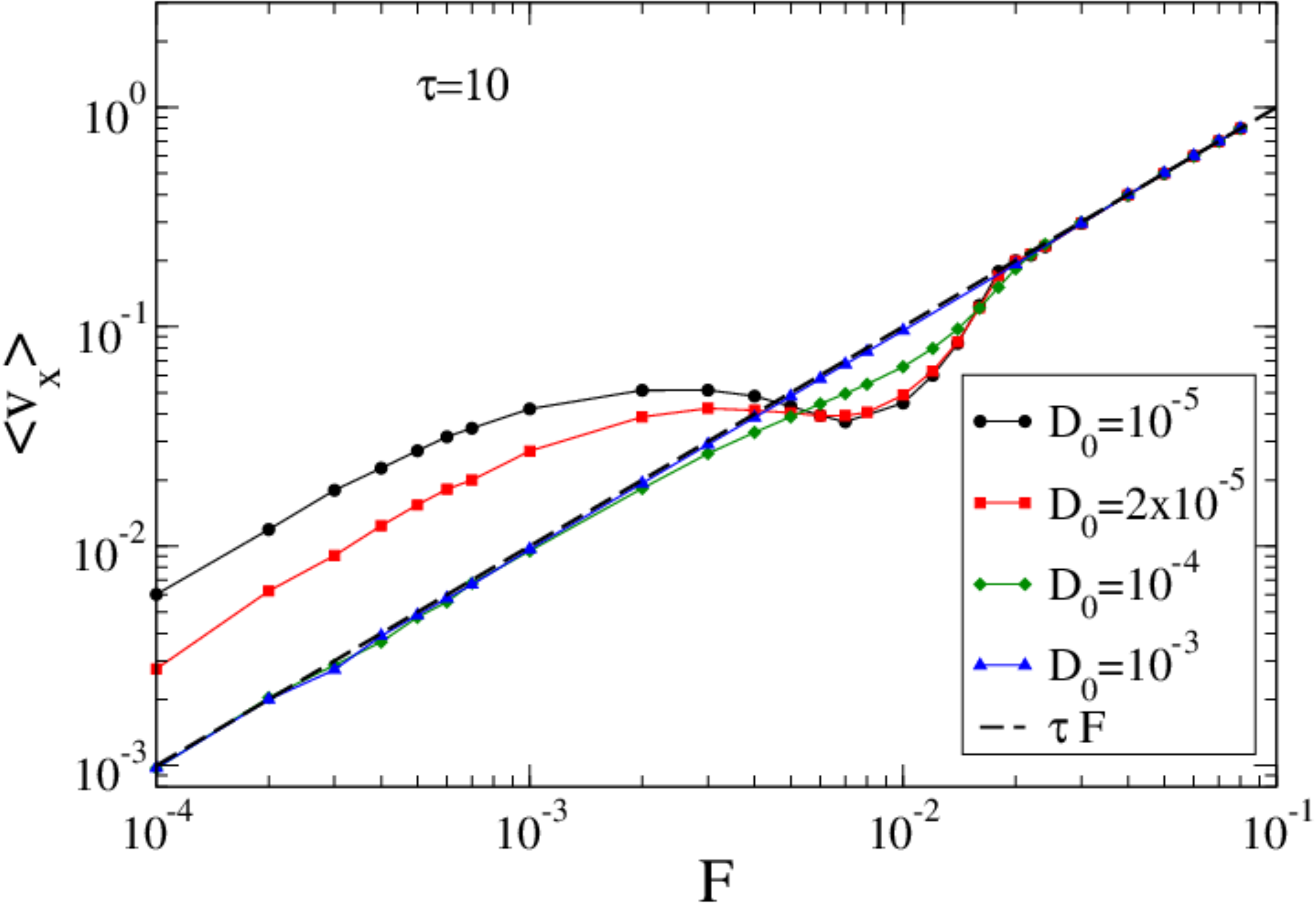}
\caption{Force-velocity relation for $\tau=10$ and several
values of $D_0$.}
\label{fig:bigTau-VF}
\end{figure}

%==================================================================
In the opposite regime, $\tau \gg\tau^*$, we find a behavior
similar to that discussed above, with NDM at small $D_0$ and monotonic
behavior for $D_0 \geq 10^{-3}$, see Fig.~\ref{fig:bigTau-VF}. However, the
drop in the velocity is much milder in this case, and the minimum is
reached for $F\approx 7\cdot 10^{-3}$. This behavior can be explained by
the fact that, since inertia is larger in this case, the action of the
velocity field is weaker and the effect of slowing down on the tracer
is reduced.
%--------------------------FIG.4--------------------------------
%----------------------------------------------------------------

%=================================================================
\subsection{Absolute negative mobility}
\begin{figure}[!th]
\includegraphics[width=.8\columnwidth,clip=true]{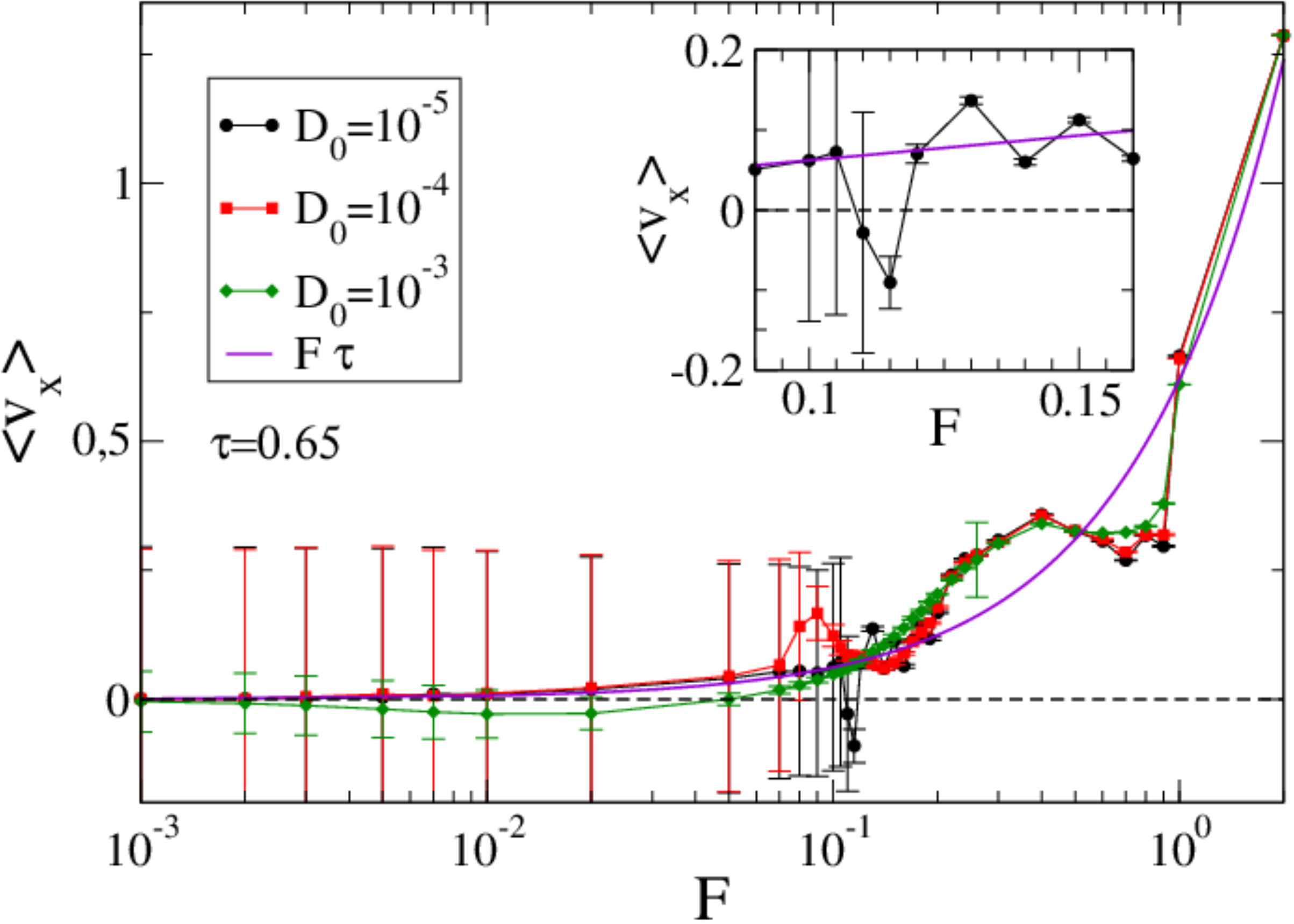}
\includegraphics[width=.8\columnwidth,clip=true]{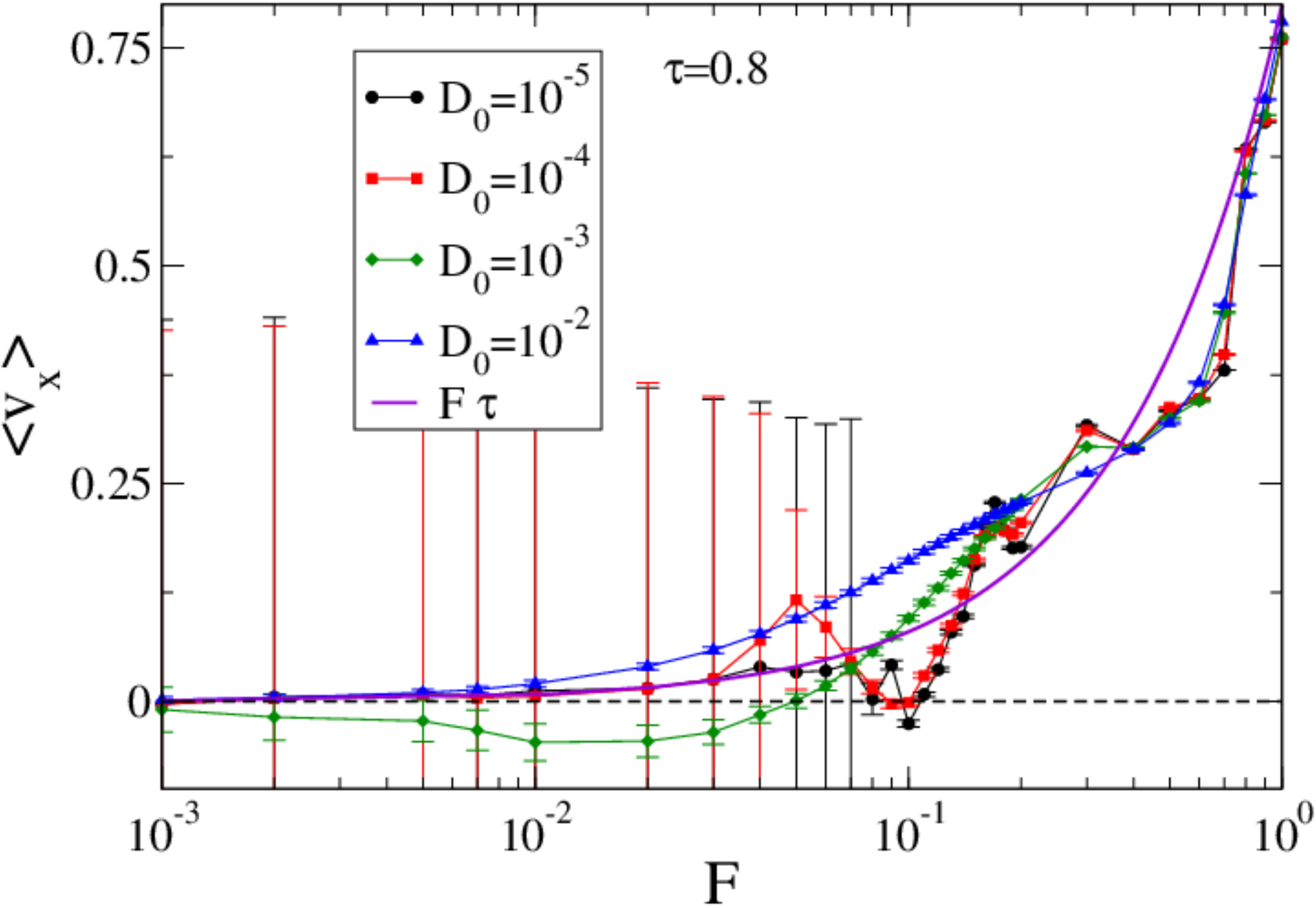}
\caption{Top panel: force-velocity relation for $\tau=0.65$ and
  several values of $D_0$. The inset shows a zoom of the ANM
  region. Bottom panel: force-velocity curve for $\tau=0.8$ and
  several values of $D_0$. The large error bars at small forces are
  due to the fact that, in those cases, the inertial particles are
  constrained along few straight trajectories, with opposite mean
  velocity, strongly dependent on the initial conditions. This effect
  produces a large variance.}
\label{fig5}
\end{figure}

%=================================================================
If the relevant time scales of the tracer and of the
underlying velocity field are comparable, $\tau \sim \tau^*$, a new
surprising effect can be observed. In Fig.~\ref{fig5} we show $\langle
v_x \rangle(F)$ for $\tau=0.65$ (top panel) and $\tau=0.8$ (bottom
panel), and different values of $D_0$.  We observe a nonlinear complex
behavior of the force-velocity relation and, in particular, we note
that the average velocity can take on negative values, $\langle
v_x\rangle<0$ within the error bars, implying an absolute negative
mobility of the tracer. This phenomenon occurs in a very narrow range
of forces for small values of $D_0$ (see black dots in
Fig.~\ref{fig5}), while for values of $D_0=10^{-3}$ negative mobility
extends at small forces, even in the linear regime (see green
diamonds). This effect is made possible by the non-equilibrium nature
of our model due to the non-gradient form of the velocity field (see
Eq.~(\ref{eq:psi})), which induces currents even in the absence of the
external force. 

In particular, our results seem to be consistent with the theoretical
analysis presented in~\cite{SER07}, for an underdamped Brownian
particle model in a one-dimensional periodic potential. Indeed, it is
expected that in some regions of the parameter space, at fixed force,
the tracer velocity follows the external force for low noise, but
changes sign upon increasing the temperature.
%----------------------------FIG.5-----------------------------------
%---------------------------------------------------------------------
A physical mechanism responsible for such a surprising effect has been
proposed in~\cite{SCPV16}, based on a careful study of the tracer
trajectories.  
This analysis showed that the motion of the tracer is
realized along preferential ``channels'', aligned downstream or
upstream with respect to the force. 
Transitions between channels are induced by the subtle interplay between 
inertia and thermal noise, analogously to the mechanism
described in~\cite{MKTLH07} for a model of a driven inertial Brownian
particle moving in a periodic potential and subject to a periodic
forcing in one dimension. 
In our case, the external force can induce a
bias in such transitions, yielding an average $\langle v_x \rangle
\neq 0$. In particular, we observed~\cite{SCPV16} that in a specific
range of forces the tracer can be biased to visit more frequently
upstream channels, slowing its motion, and leading to NDM or even to
ANM. An analogous mechanism has been described in \cite{SER07b}.

%=====================================================================
\subsection{Phase chart}
\begin{figure}[!th]
\includegraphics[width=1.\columnwidth,clip=true]{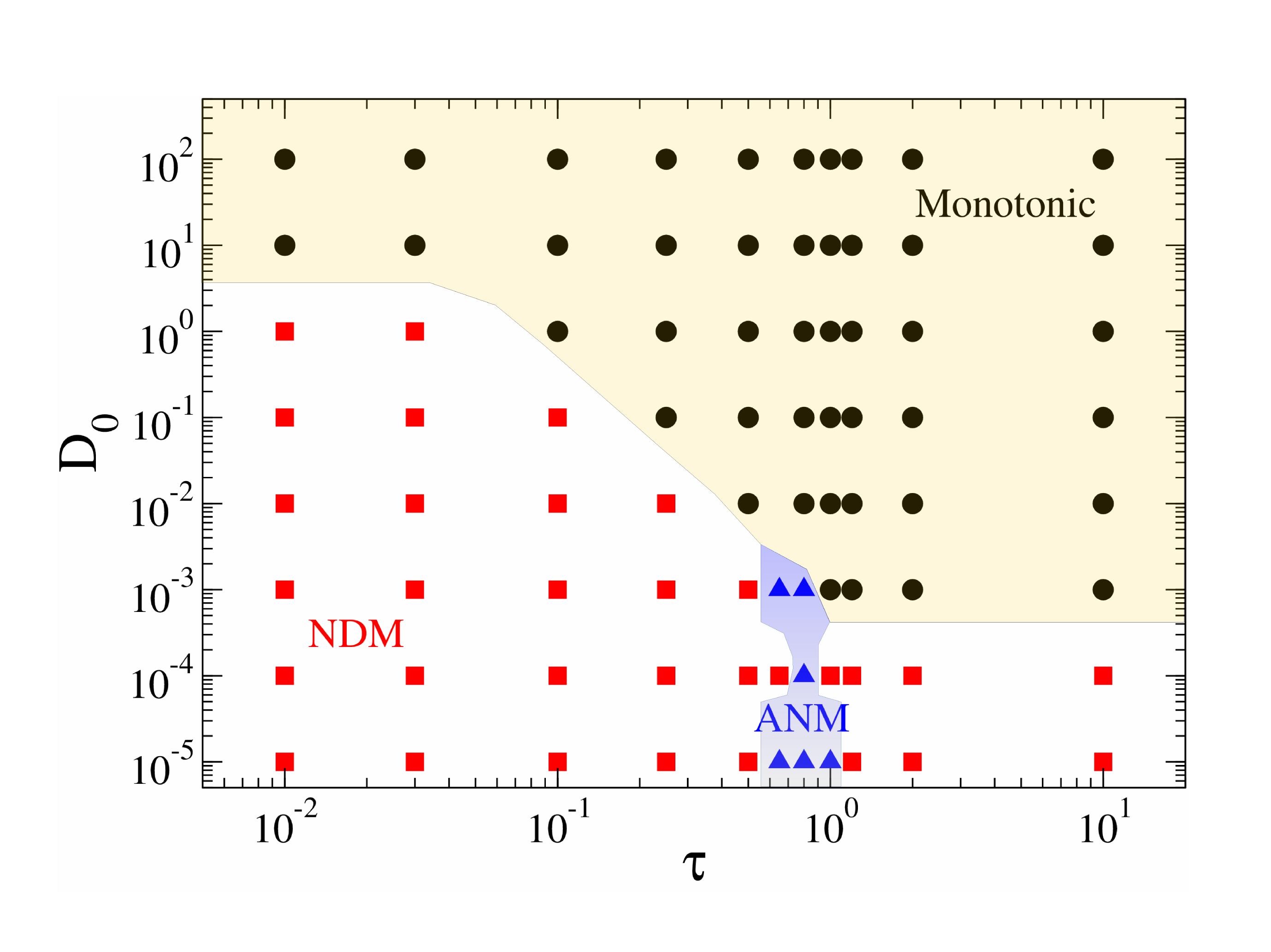}
\caption{Phase chart in the parameter space of the model. Black dots
  identify the region where monotonic behavior of the force-velocity
  relation is observed, red squares regions where NDM takes
  place, and blue triangles regions where ANM occurs.}
\label{fig:phase}
\end{figure}

%=====================================================================
Our extensive numerical study of the model is summarized in the phase
chart reported in Fig.~\ref{fig:phase}. We identified three
``phases'', corresponding to simple monotonic behavior (black dots),
NDM (red squares) and ANM (blue triangles), for the velocity-force
relation, in the parameter space $(\tau,D_0)$. For each couple of
parameters, we performed numerical simulations studying the curve
$\langle v_x\rangle(F)$ and focusing on the regions where non-linear
behaviors occur.
%-----------------------------FIG.6-------------------------------------
%------------------------------------------------------------------------
As expected, for large values of the microscopic diffusivity $D_0$,
the system exhibits a simple behavior, because noise dominates over
the effect of the underlying velocity field, and a monotonic behavior
is observed. The same happens for large values of $\tau$, when again
the underlying field can be neglected.  NDM seems a typical phenomenon
for small values of $\tau$ and $D_0$, where the different driving
mechanisms acting on the tracer are comparable and coupled, leading to
non-monotonic behaviors. Our study also unveils a narrow region where
ANM can be observed, which occurs for values of $\tau\sim\tau^*$,
where $\tau^*$ is the typical time scale of the underlying velocity
field. For $\tau\to 0$, the occurrence of ANM can be excluded due to the no-go theorem discussed in~\cite{SER07}.

%%%%%%%%%%%%%%%%%%%%%%%%%%%%%%%%%%%%%%%%%%%%%%%%%%%%%%%%%%%%%%%%%%%%%%%
\section{Conclusions}
%%%%%%%%%%%%%%%%%%%%%%%%%%%%%%%%%%%%%%%%%%%%%%%%%%%%%%%%%%%%%%%%%%%%%%%
We have investigated a model for the dynamics of an inertial tracer
advected by a laminar velocity field, under the action of an external
force and subject to thermal noise. This model can be useful to
describe the transport properties of small particles (e.g. soil dust,
man-made pollutants or swimming microorganisms~\cite{CGCB17})
dispersed in fluids in different contexts, such as aerosol
sedimentation, plankton concentration, or gravitational settling, with
application in several areas of engineering, oceanography or
meteorology.

We focused on the force-velocity relation of the tracer in the
nonlinear forcing regime. The system shows a very rich phenomenology,
featuring negative differential and absolute mobility, as summarized
by the phase chart in the model parameter space. The emergence of
these effects is due to the subtle coupling between the velocity field
dynamics and the inertia of the tracer, and crucially depends on the
amplitude of the microscopic thermal noise and on the time scale ratio
between the tracer Stokes time and the characteristic time of the
fluid. The presence of two non-equilibrium sources in the system,
namely the non-gradient force and the external bias, and its combined
action, make the model very rich: beyond the known trapping effects
which result in a reduction of the tracer velocity upon increasing the
applied force, in our model we also observe regions of negative
mobility, where the particle travels against the external force.  This
phenomenon demands for experimental investigation in real systems.

\section*{Author contribution statement}
All authors contributed equally to the paper.

%
% BibTeX users please use
% \bibliographystyle{}
% \bibliography{}

\end{document}